\documentclass[universe,article,accept,pdftex,moreauthors]{Definitions/mdpi} 

\newcommand{\diff}{{\mathrm d}}
\newcommand{\sa}{{$a$}}
\newcommand{\lcdm}{{$\Lambda \rm CDM$}}
\newcommand{\hmpc}{{$h^{-1}\rm Mpc$}}

\firstpage{1} 
\pubvolume{11}
\issuenum{7}
\articlenumber{212}
\pubyear{2025}
\copyrightyear{2025}
\externaleditor{Hermano Velten} % More than 1 editor, please add `` and '' before the last editor name
\datereceived{25 April 2025} 
\daterevised{21 June 2025} % Comment out if no revised date
\dateaccepted{23 June 2025} 
\datepublished{26 June 2025} 
\hreflink{https://doi.org/10.3390/ universe11070212} % If needed use \linebreak
\def\hl{} 
\def\highlighting{} 

\Title{Cosmological Simulations with Massive Neutrinos: Efficiency and Accuracy}

\TitleCitation{Cosmological Simulations with Massive Neutrinos: Efficiency and Accuracy}

\Author{\hl{Bing-Hang Chen} %MDPI: 1. Please carefully check the accuracy of names and affiliations. 2. For Chinese authors, both formats of “Xiaoming Wang” and “Xiao-Ming Wang” are acceptable; for authors from mainland China, the format “Xiaoming Wang” is preferred. The name format should be as uniform as possible, please check all highlight names
 $^{1, \dagger }$\orcidA{},  \hl{Jun-Jie Zhao} $^{2, \dagger }$\orcidC{}, \hl{Hao-Ran Yu} $^{1,*}$\orcidB{}, Yu Liu $^{3}$\orcidF{}, \hl{Jian-Hua He} $^{4}$\orcidD{} and \hl{Yipeng Jing} $^{2}$\orcidE{}}

\AuthorNames{Bing-Hang Chen, Jun-Jie Zhao, Hao-Ran Yu, Yu Liu, Jian-Hua He and Yipeng Jing}

\isAPAStyle{%
       \AuthorCitation{Chen,B.; Zhao,J.; Yu,H.; Liu,Y.; He,J.;Jing,Y.}
         }{%
        \isChicagoStyle{%
        \AuthorCitation{Chen,Bing-Hang; Zhao,Jun-Jie; Yu,Hao-Ran; Liu,Yu; He,Jian-Hua; Jing,Yipeng}
        }{
        \AuthorCitation{Chen, B.-H.; Zhao, J.-J.; Yu, H.-R.; Liu, Y.; He, J.-H.; Jing, Y.}
        }
}

\address{%
$^{1}$ \textls[-15]{\quad Department of Astronomy, Xiamen University, Xiamen 361005, China} %MDPI: We added the email addresses here according to those submitted  at susy.mdpi.com. Please confirm.
\\
$^{2}$ \quad \textls[-15]{State Key Laboratory of Dark Matter Physics, Tsung-Dao Lee Institute \& School of Physics and Astronomy, Shanghai Jiao Tong University, Shanghai 201210, China}\\
$^{3}$ \quad Department of Astronomy, Tsinghua University, Beijing 100084, China\\
$^{4}$ \quad School of Astronomy and Space Science, Nanjing University, Nanjing 210093, China
}

\corres{Correspondence: haoran@xmu.edu.cn}

\firstnote{These authors contributed equally to this work.}

\abstract{
Constraining neutrino mass through cosmological observations relies on precise simulations to calibrate their effects on large scale structure, while these simulations must overcome computational challenges like dealing with large velocity dispersions and small intrinsic neutrino perturbations. We present an efficient $N$-body implementation with semi-linear neutrino mass response which gives accurate power spectra and halo statistics. We explore the necessity of correcting the expansion history caused by massive neutrinos and the transition between relativistic and non-relativistic components. The above method of including neutrino masses is built into the memory-, scalability-, and precision-optimized parallel $N$-body simulation code \highlighting{{\tt CUBE}} %MDPI: Please check if need keep this font. please check it in whole text
 2.0. Through a suite of neutrino simulations, we precisely quantify the neutrino mass effects on the nonlinear matter power spectra and halo statistics. 
}

\keyword{cosmology\highlighting{;} %MDPI: We changed ``:'' to ``;'', please confirm 
 neutrinos\highlighting{;} %MDPI: We changed ``–'' to ``;'', please confirm 
 dark matter\highlighting{;} %MDPI: We changed ``–'' to ``;'', please confirm 
 large scale structure of Universe; \mbox{methods}\highlighting{;} %MDPI: We changed ``:'' to ``;'', please confirm 
  numerical} 

\begin{document}

\section{Introduction}
The {\lcdm} model is the current standard model of cosmology. Increasingly large and precise observational results have largely confirmed its accuracy, such as cosmic microwave background (CMB), large scale structure and gravitational lensing. Future observation projects lead us into an era of higher precision in order to detect potential deviations from a pure cold dark matter (CDM) model, such as nonzero neutrino masses, primordial non-Gaussianities, and variations of gravity. Among these, the neutrino mass is proven to be nonzero by particle physics and neutrino mass effects on cosmology are guaranteed to exist. 
In laboratory experiments, the neutrino mass-squared differences are determined by measuring the atmospheric and solar neutrino oscillations \cite{ParticelGlobal}, indicating that the total mass of neutrinos is nonzero, with a lower bound of $M_\nu \equiv \Sigma m_{\nu.i} \gtrsim 0.06 \, \mathrm{eV}$, although they cannot determine the absolute neutrino mass or the mass hierarchy. The tritium $\beta$-decay experiment, like KATRIN, gives an upper limit of $0.8 \, \mathrm{eV}$ at a 90\% confidence level \cite{AkeNat.Phys.2022}.  On the other hand, cosmological observations provide an upper limit on the total neutrino mass $M_\nu \lesssim 0.09 \, \mathrm{eV}$ \cite{DiPhys.Rev.D2021}. The absence of lower bound constraints on neutrino mass in cosmology implies that the absolute mass of neutrinos remains unmeasured. Detecting the cosmological effects of neutrino mass and precisely constraining its value are key objectives of next-generation galaxy surveys.

From the theoretical side, neutrinos affect cosmic evolution through two primary mechanisms. In the early Universe, relativistic neutrinos have high velocity dispersion, behaving  as radiation. They modulate the expansion history by changing the matter--radiation equality \cite{LESGOURGUES_2006}, altering the acoustic and damping angular scales of CMB \cite{BasPhys.Rev.D2004}. In the late Universe, neutrinos cool down and become non-relativistic, while their thermal velocities are still higher than those of CDM; this free-streaming mechanism suppresses their gravitational clustering and thus damps the small-scale matter power spectrum and the high multipoles of the CMB anisotropy, and modulates the halo mass function (HMF) and halo bias \cite{Euclid,MTNG}. 
Apart from traditional methods, various novel methods have been proposed to constrain the lower bound of neutrino mass, for example, a dipole asymmetry in the cross-power spectrum due to the velocity field between neutrinos and CDM \cite{dipole}, wakes formed as neutrinos passing through dark matter halos \cite{Wakes}, misalignments in angular momentum between neutrino and CDM halos due to differing tidal fields \cite{Torque}, and differential neutrino condensation between neutrino-rich and neutrino-poor halos \cite{Condensation}.  Neutrino perturbations couple to nonlinear CDM clustering, and thus, the above effects are also nonlinear, so we must use simulations to accurately calibrate these effects. Additionally, a combination of variations of cosmological parameters may also mimic neutrino mass effects, so a large number of simulations spanning a high-dimensional parameter space will also be needed.
To achieve this and meet the scientific requirements for next-generation surveys, such as CSST \cite{Han2025}, we need a method to simulate the effects of massive neutrinos that balances accuracy and efficiency.

Efficiently \textls[-15]{simulating neutrinos is challenging due to their high thermal} \mbox{velocities \cite{Hydrodynamics, ZenJCosmolAstropartP2019}}. Particle-based approaches partition neutrino phase space into $N$-body particles; however, they suffer from prohibitive shot noise on small scales due to the neutrino thermal velocities that contaminate small-scale signals. Increasing the number of neutrino particles to suppress this noise is computationally expensive \cite{TianNu}. Alternative methods have been developed to simulate neutrino effects.
The grid-based method computes only CDM particles and incorporates neutrino effects by adding an independently evolved linear neutrino density field in Fourier space, which is a fully linear approach \cite{BraJ.Cosmol.Astropart.Phys.2009}. 
The semi-linear method improves the linear approach by taking into account the nonlinear evolution history of matter density perturbations to estimate neutrino density perturbations \cite{AB13}. It is more accurate than the fully linear method and is still efficient \cite{MassiveNuS, MTNG}. %But it may  be imprecise on small scales due to the lack of out-of-phase information between neutrino and CDM.
A recent extension, the SuperEasy method \cite{Chen:2021}, further simplifies the linear response implementation by introducing a closed-form rational function to approximate neutrino clustering, achieving sub-percent accuracy with minimal computational cost.
The hybrid method treats neutrinos as linear fields in the early Universe, computes their evolution on grids, and converts some of the slower neutrinos into particles at low redshift during simulation, thereby capturing their nonlinear clustering properties more effectively \cite{B18, MTNG}. 
The fluid method models different neutrino species as fluids rather than particles, solving the Boltzmann equations directly on a \mbox{grid \cite{Hydrodynamics, Liu2020, Multi-Fluid}}. While this method can accurately reproduce nonlinear clustering, its numerical complexity often results in instability and artificial numerical artifacts in simulations. 
The Newtonian motion gauge approach \cite{FidJ_2019,Partmann_2020, Heuschling_2022} provides a simple method to include massive neutrino effects in N-body simulation. It incorporates the additional potential contribution of neutrinos via modification of the dark matter initial conditions and by employing a dynamically
evolving coordinate system \cite{Euclid}, avoiding direct neutrino simulations. It can also capture the full impact of linear neutrino perturbations on the nonlinear evolution of CDM particles in N-body simulation.

% In order to generate a series of high precision simulations with neutrino mass effect for the scientific preparation for next generation surveys, such as CSST \cite{Han2025}, we need to banlence between accuracy and  efficiency. 
Given the fact that no neutrino effects have been discovered observationally from large scale structure, we tend to focus on lower-order effects. The semi-linear method captures most of the neutrino-CDM nonlinear coupling, although it neglects out-of-phase correlations. Despite this, it can accurately model key nonlinear neutrino effects, such as HMF modulation and neutrino torque. We have incorporated a revised version of the semi-linear neutrino method into {\tt CUBE} 2.0, a parallel, high-precision, and memory-efficient $N$-body algorithm. Our implementation enables efficient high-resolution large scale structure simulations with varying neutrino masses and hierarchies. In Section \ref{sec2}, we briefly introduce the code and relevant algorithm. In Section \ref{sec3}, we analyze the accuracy of the simulation.  Discussions are presented in Section \ref{sec4}.

%%%%%%%%%%%%%%%%%%%%%%%%%%%%%%%%%%%%%%%%%%
\section{Methods}\label{sec2}

\subsection{CDM Evolution}
We use {\tt CUBE} 2.0 as the basic structure to evolve the overall large scale structure, and the neutrino module is an add-on functional. {\tt CUBE} 2.0 is an efficient, parallel cosmological $N$-body simulation code based on the Particle-Mesh-Particle-Particle ($\rm P^3M$) algorithm. The preceding version, {\tt CUBE} \cite{CUBE}, optimized the memory usage by the fixed point structure (FPS) storage, reducing the memory footprint as low as 6 bytes per particle. {\tt CUBE} 2.0 implements an adaptive innermost grid (PM3) to complement the original two-layer PM (PM1/PM2) framework to substantially alleviate the particle-particle (PP) computation. The resolution of the adaptive grid matches the local density, ensuring that the total computation time of both PP and PM3 is minimized. Furthermore, by employing the optimized Green's \mbox{function \cite{XuApJ2021}} for gravitational calculations, the code achieves maximum relative errors below $5\%$ (with mean errors below $2\%$) in pairwise gravitational interactions \mbox{between particles}.

The advantages of our cosmological simulations lie primarily in their highly memory-efficient storage of particle information. On average, our method requires only 20 bytes per particle---a significant reduction compared to the widely used Gadget2, which consumes approximately 80 bytes per particle \cite{Springel2005}. This represents a several-fold optimization in memory usage for dark matter particle storage \cite{Gadget4}.

In terms of computational efficiency, {\tt CUBE} 2.0 has been validated for large scale simulations (over 16,000 cores in 512 nodes) [Yu et al. in prep.], demonstrating the robustness of its three-layer parallel structure. It achieves strong and weak scalability exceeding $90\%$, while maintaining its advantage in the memory economy.

\subsection{Semi-Linear Neutrino Modeling}%Incorporating Massive Neutrinos}

% \subsubsection{Relativistic to non-relativistic transition}
The neutrino module is built on the above CDM evolution. We consider the following components of the universe---the cosmological constant ($\Omega_\Lambda$), radiation ($\Omega_\gamma$), CDM ($\Omega_{\rm c}$), baryonic matter ($\Omega_{\rm b}$) and neutrinos ($\Omega_\nu$). Since we do not distinguish CDM and baryons in this work, we can safely absorb baryons into CDM, and they are collectively referred to as ``cold matter'' ($\Omega_{\rm cb} = \Omega_{\rm c} + \Omega_{\rm b}$). Neutrinos can be divided into relativistic ($\Omega_\nu^{\rm r}$) and non-relativistic ($\Omega_\nu^{\rm nr}$) components, with cosmic time dependency. Therefore, the total matter component is ($\Omega_{\rm M} = \Omega_{\rm cb} + \Omega_\nu^{\rm nr}$).

To better understand the two components of neutrinos, we describe neutrinos as a fluid with an effective pressure $p_\nu$. The equation of state is given by $p_\nu = \omega_\nu \rho_\nu$, where $\omega_\nu(a)$ is a function that decreases with \sa. Relativistic neutrinos can thus be \hl{described as} %MDPI: Please carefully check variable formatting (italic, bold, subscript, uppercase, etc.) throughout the manuscript to ensure the formatting is consistent and revise if needed. 

\begin{equation}
    \Omega_{\nu}^{\rm r}(a) = 3 w_{\nu}(a) \Omega_{\nu}(a).
\end{equation}
\hl{The evolution} %MDPI: Please check if need add indent. The same for the following highlights.
%No, its't need to add indent
 of neutrinos satisfies $\rho_\nu(a) \propto a^{-3[1+w_\nu(a)]}$, or equivalently,

\begin{equation}
\begin{aligned}
3(1 + w_\nu) = -\frac{\mathrm{d} \ln \rho_\nu}{\mathrm{d} \ln a} = 4 - y \frac{\mathrm{d} \ln \mathcal{F}(y)}{\mathrm{d} y},
\end{aligned}
\end{equation}
where $y = aM_\nu/\left(\Gamma_\nu N_\nu k_{\rm B} T_{\gamma,0}\right)$,  and $T_{\nu,0}$, $\Gamma_\nu \equiv T_{\nu,0}/T_{\gamma,0}$ is the neutrino temperature and the neutrino-to-photon temperature ratio today. $N_\nu$ is the number of massive neutrinos, $k_{\rm B}$ is the Boltzmann constant, and
\begin{equation}
\mathcal{F}(y) \equiv \int_0^\infty \frac{x^2 \sqrt{x^2 + y^2}}{1 + e^x} \, \mathrm{d} x.
\end{equation}
\hl{We also} define the fraction of non-relativistic neutrinos
%$f^{\rm nr} $, which is
\begin{equation}
\label{fnr}
\begin{aligned}
f^{\rm nr} \equiv \frac{\Omega_{\nu}^{\rm nr}}{\Omega_{\nu}^{\rm r}} = (1 - 3\omega_\nu)= \frac{y}{\mathcal{F}(y)} \frac{\mathrm{d} \mathcal{F}(y)}{\mathrm{d} y}.
\end{aligned}
\end{equation}

The semi-linear response method models neutrino effects by correcting the CDM density (no neutrinos) field into the total matter density \hl{field} %MDPI: Please check if need keep all bold format in equation variables.
 $\delta_{\rm {M}}(\boldsymbol{k},\tau)=F_\nu(k)\delta_{\rm cb}(\boldsymbol{k},\tau)$, where the correction function $F(k)$ is defined as
\begin{equation}
\label{cf}
F_\nu(k) \equiv \sqrt{P_{\rm M}/P_{\rm cb}}.
\end{equation}
\hl{This} enables CDM particles to evolve under the gravitational potential of the total matter field.  Incorporating the non-relativistic fraction $f^{\rm nr}$ from Equation \eqref{fnr}, the correction function resolves as
\begin{equation}
F_\nu(k,a) = \frac{(1-f_\nu)P_{\rm cb}^{1/2}(k,a) + f_\nu f^{\rm nr}(a)P_{\nu}^{1/2}(k,a)}{P_{\rm cb}^{1/2}(k,a)}.
\end{equation}
\hl{Here} $P_{\rm cb}(k)$ is the CDM power spectrum calculated in simulation (see Section~\ref{spk}). $P_{\nu}(k)$ is the neutrino power spectrum using equation form Equation (63) of \cite{AB13}
%\hl{Equation (63),} %MDPI: There is no eq.63 in thsi paper, please check if it is equation in reference
% I have checked and confirmed that this equation exists in this article
 derived from the integral of the total matter power spectrum evolution.

\begin{figure}[H]

\begin{adjustwidth}{-\extralength}{0cm}
\centering %% If there is a figure in wide page, please release command \centering
 \includegraphics[width=0.95\linewidth]{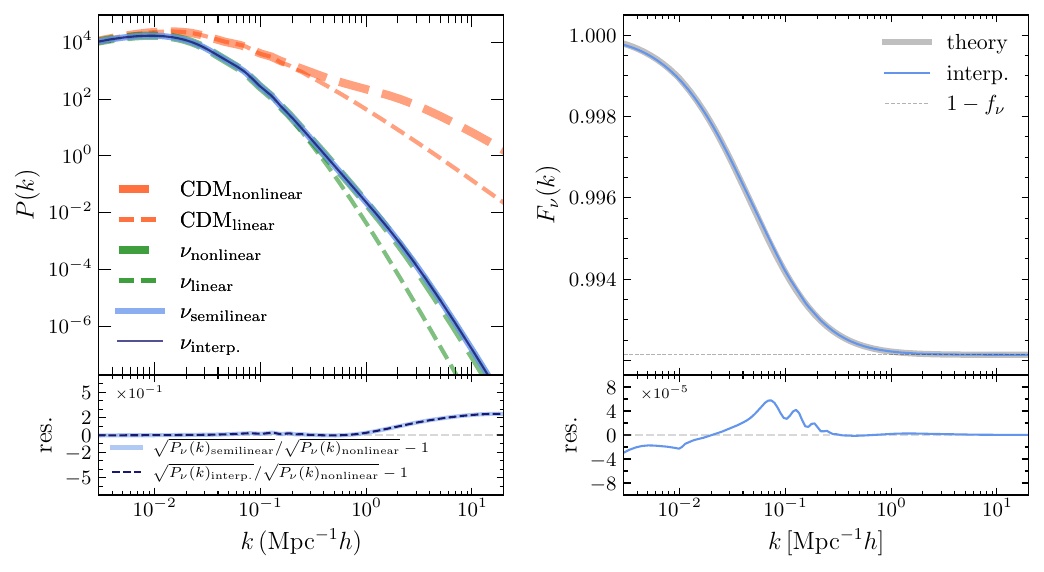}
\end{adjustwidth}
    \caption{(\textbf{\hl{Left} %MDPI: We moved all the figures after the first citation, please confirm. Please note that changes to the position/size of figures or tables may occur during the production stage. Please ensure the different color of lines are corresponding with them explanation.
    panel}): Theoretically predicted CDM (orange) and neutrino (green) power spectra are represented by dashed lines, while the power spectrum predicted by the semi-linear method is shown with solid lines. The thin solid line represents the predicted power spectrum from interpolation. The lower panel displays the residuals between the two predictions. (\textbf{Right panel}): Massive neutrino correction function from  theoretical predictions (gray) and semi-linear simulations (blue) at $z=0$, and residual plots showing difference in lower panel.}
    \label{fig:interp}
\end{figure}

To test the accuracy of the finite integral, we estimate the neutrino power spectrum using the semi-linear method, derived from the theoretical nonlinear matter power spectra generated by {\tt CAMB}, and test the interpolation strategy (labeled as “interp.”). In Figure~\ref{fig:interp}, we present the residuals of the square root of the power spectrum (rather than the power spectrum itself, as they are more closely related to errors in the correction function), which agree with the predictions of the nonlinear theory on scales $k < 1 $, while significant deviations emerge on small scales. However, these small-scale discrepancies are negligible, as demonstrated by the redshift $z=0$ correction function; for $k > 1 $, the correction function drops to $1-f_\nu$, indicating that neutrinos behave nearly homogeneously on this scale. This ensures that, despite slight deviations in the power spectra estimates, the overall correction function aligns perfectly with theoretical predictions, with the relative error controlled below $10^{-4}$. 
% For consistency, we introduce an additional $(1-f_\nu)^2$ factor to the particle-particle (PP) interaction forces, effectively accounting for neutrino effects at these scales.
Since the wavenumber at the particle-particle (PP) scale is evidently much greater than 1, we assume that neutrinos are completely homogeneous at this scale. Accordingly, the effective mass contributing to the gravitational interaction between particles should be scaled by a factor of $(1 - f_\nu)$, leading to a correction factor of $(1 - f_\nu)^2$ for the PP interaction force.

% \subsubsection{Expansion History}

The expansion history of the universe, which couples with the effects of neutrinos, influences the clustering of matter, making it necessary to use an accurate expansion history. {\tt CUBE} solves the expansion history during simulation steps (on-the-fly computation) and introduces subtle discrepancies in the expansion history under varying simulation resolutions and precisions. Although these variations diminish with reduced time-step size and remain acceptable in pure CDM simulations, they become problematic for massive neutrino simulations that require higher accuracy and stability. To address this, we implement a precomputed expansion history obtained by solving the Friedmann equation:

\begin{equation}
H(a) = H_0 \left[
\Omega_\Lambda + \Omega_{\rm cb} a^{-3} + \Omega_\gamma a^{-4} +  
\frac{15}{\pi^4} \Gamma_{\nu}^{4} N_{\nu} \Omega_{\gamma,0} \mathcal{F}( y) a^{-4}
\right]^{1/2},
\end{equation}
where the final term accounts for the neutrino component \cite{IC}.

For convenience, we set the present superconformal time $s(a) \equiv \int_0^{t(a)}a^{-2} {\rm d}t$ to zero, and we integrate backward in time to obtain the expansion history at higher redshifts:

\begin{equation}
a(s) = \int_0^{s} a^{3} H(a) (-\diff s).
\end{equation}

\begin{figure}[H]
   
    \includegraphics[width=0.7\linewidth]{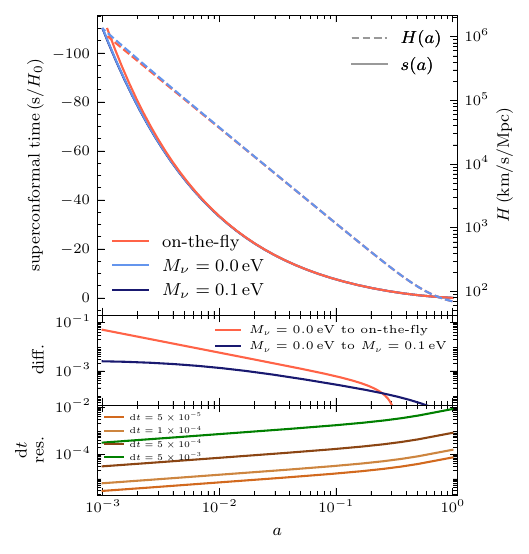}
    \caption{\hl{Expansion} %MDPI: Please ensure the different color of lines are corresponding with them explanation.  Please check if the 'a' at the bottom of the image is the subgraph number. If it is, please remove it
 history (solid line) and Hubble expansion rate (dashed line), as functions of scale factor $a$,  computed using the on-the-fly method (red). Different neutrino masses (blue) cause a $1\%$ variation in the expansion history. The bottom panel shows the impact of the integration time-step size on accuracy.}
    \label{fig:EH}
\end{figure}

We show the differences in the expansion history and Hubble expansion rate between the on-the-fly algorithm and the new algorithm in Figure~\ref{fig:EH}. The error between the two algorithms reaches a maximum of around $10\%$, which can lead to significant differences in large scale structure evolution. On the other hand, the variations in the expansion history for different neutrino masses are on the order of $\lesssim 1\%$, meaning that the integration error must be controlled below $10^{-4}$. Therefore, choosing an integration step size of $5 \times 10^{-4}$ is sufficient to meet this requirement.

\subsection{Power Spectrum Estimation}
\label{spk}
The semi-linear method requires integrating the matter power spectrum to obtain the neutrino power spectrum. To account for effects such as cosmic variance, we need to compute the matter power spectrum from simulations rather than using a theoretical one. However, calculating the global power spectrum at the same resolution as the simulation is time- and memory-consuming, and can be comparable to executing a simulation step at low redshift. Therefore, computing the global power spectrum at each step is not practical. Reducing the resolution of the global power spectrum would improve computational efficiency, but would also result in the loss of much nonlinear information. To address this, we calculate the global power spectrum at specific redshifts $z_{\rm p}$ and use interpolation and extrapolation to estimate the power spectrum at other redshifts. Our analysis compares various $z_{\rm p}$ selection schemes and interpolation algorithms, revealing that quadratic interpolation maintains the relative error of the matter power spectra within $0.2\%$ while achieving neutrino power spectrum accuracy at the $10^{-4}$ level relative to non-interpolated methods. This approach yields correction functions with residual errors below $10^{-5}$, satisfying the precision requirements for most cosmological studies.

\begin{figure}[H]
   
    \includegraphics[width=0.7\linewidth]{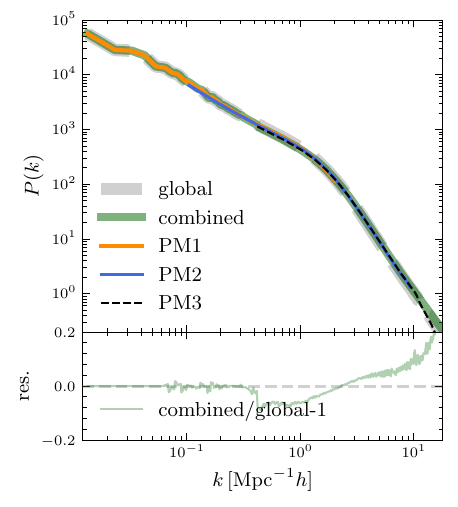}
    \caption{\hl{Multi-scale} %MDPI: Please ensure the different color of lines are corresponding with them explanation.
  power spectrum combination (thick green curve), compared with the power calculated from the full-resolution algorithm (thick gray curve). Thinner curves show the power spectra calculated by different PM grids. The bottom panel shows the residual.}
    \label{fig:Merge}
\end{figure}

To control the memory consumption in power spectrum calculations during simulations, we develop a power spectrum combining approach by extending Welch’s \mbox{method \cite{Welch}} to three dimensions and integrating it with buffer designs from Particle-Mesh (PM) algorithms. This involved the creation of novel 3D window functions that maintain uniform weighting across all data points with minimized variance. Our approach accelerates the process by a factor of 10 compared to full-resolution global calculations while maintaining equivalent accuracy (Figure~\ref{fig:Merge}), crucially avoiding additional memory overhead since density field computations are intrinsic to PM algorithms. Our combining power spectrum implementation proceeds as follows.
\begin{enumerate}
\item Compute the low-resolution global power spectrum $P_{\rm PM1}(k)$ at the PM1 grid points.
\item Calculate tiered spectra $P_{{\rm PM}x}(k)$ at standard and high resolutions (PM2/PM3 grids) through all subvolumes:
\begin{enumerate}
    \item Apply the window function to the density field within the subvolume.
    \item Calculate $P_{{\rm{raw}} ,i}(k)$, then perform a noise correction \cite{jing} to obtain $P_{{\rm denoised},i}(k)$.
    \item Average all $P_{{\rm denoised},i}(k)$, then apply the alias correction \cite{jing} to obtain $P_{{\rm PM}x}(k)$. 
    \item Automatically apply a cut-off on the power spectrum to account for grid effects.
\end{enumerate}
\item Combine the three levels of $P_{{\rm PM}x}(k$) based on the sampling count and obtain $P_{\rm combine}(k)$.
\end{enumerate}

\vspace{-3pt}

Due to the fact that semi-linear corrections require the application of correction functions to the grid, the $k$-modes of the power spectrum cannot perfectly match all $k$-modes of the density field, especially when smoothing the $k$-modes via binning. We choose to retain all actual $k$-modes below $k_{\rm grid}<6$ to prevent interpolation errors in sparsely sampled low-frequency regions, implementing interpolation only at higher frequencies where spectral sampling density permits reliable estimation.

%%%%%%%%%%%%%%%%%%%%%%%%%%%%%%%%%%%%%%%%%%
\section{Simulations}\label{sec3}
% \subsubsection{Simulation}
The above neutrino module is built into the parallel $N$-body code {\tt CUBE} 2.0. By using it, we run a series of simulations with different resolutions, box sizes, and neutrino masses (see Table \ref{tab:sim} for specific parameters). In tuning neutrino masses, we set the density parameters of total matter and baryons to $\Omega_{\rm M} = 0.279$ and $\Omega_{\rm b} = 0.0436$, respectively. In the simulation with massless neutrinos, $\Omega_{\rm M}  = \Omega_{\rm b} + \Omega_{\rm c}$. In the simulation with massive neutrinos, $\Omega_{\rm M}  = \Omega_{\rm b} + \Omega_{\rm c} + \Omega_{\nu}$, where a portion of the dark matter is replaced by neutrinos. The dimensionless Hubble constant $\textit{h}$ is set to $0.7$. This section presents the data results from these simulations, starting with an analysis of the matter clustering and then examining the halo statistics.

\begin{table}[H]
\caption{\hl{Neutrino} %MDPI: Please confirm the border change. We removed empty rows, please confirm
 simulation configurations.}\label{tab:sim}
\setlength{\tabcolsep}{9mm}
\begin{tabular}{cccc}
\toprule

\textbf{Simulation} & \textbf{Box} \boldmath{[\hmpc]}        & \boldmath{$N_{\rm CDM}$} & \boldmath{$\sum m_\nu$ \textbf{[}$\rm eV $\textbf{]}} \\
% &  [\hmpc]  &           &  [$\rm eV $]   \\

\midrule

     &      &          & 0.05 \\ 

600,512  & 600  & $512^3$  & 0.10 \\
&      &          & 0.15 \\
\midrule
300,512  & 300  & $512^3$  & 0.10 \\

1200,512 & 1200 & $512^3$  & 0.10 \\

600,1024& 300  & $1024^3$ & 0.10 \\

\bottomrule

\end{tabular}

\end{table}

\subsection{Matter Clustering} 
\label{MC}

\begin{figure}[H]
  
    \includegraphics[width=0.7\linewidth]{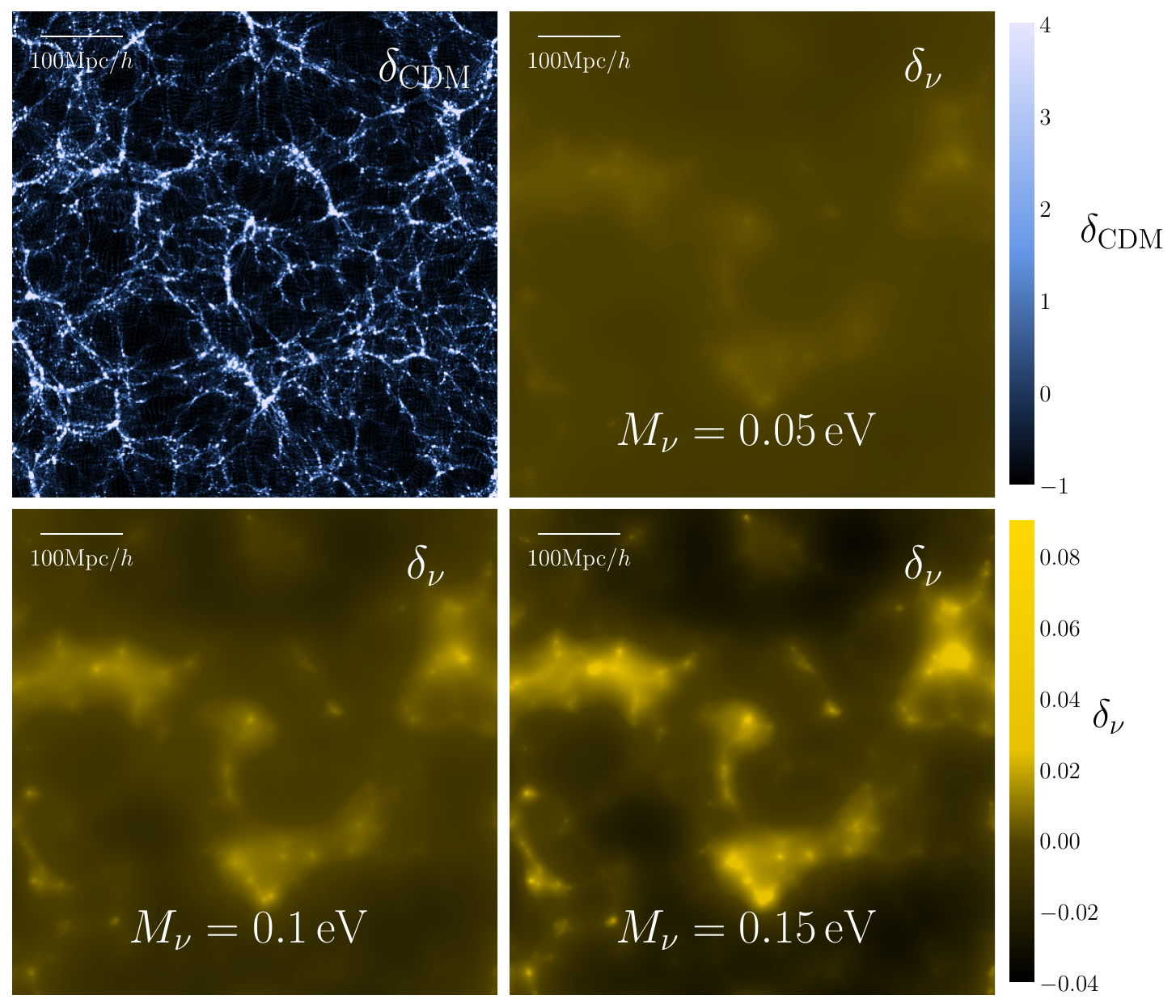}
    \caption{CDM (top left panel) and neutrino overdensity projections with a thickness of 1.2 {\hmpc}. The CDM overdensity is obtained from the simulation whose neutrino mass set to zero, while it is indistinguishable from the ones with neutrino mass. The rest of the panels show neutrino overdensities from respective simulations.}
    \label{fig:rho}
\end{figure}

In Figure~\ref{fig:rho}, we visualize the results of simulations by CDM and neutrino overdensities. The top left panel shows the CDM overdensity from a slice of 1.2 {\hmpc} thickness, taken from the massless neutrino simulation, showing the obvious cosmic web structure. Although there would be trace differences if it were taken from non-zero-neutrino-mass simulations, they are indistinguishable by eye, so we plot only one of them. The rest three panels show neutrino overdensities from the simulation with sum of neutrino masses set to  $0.05 \, \mathrm{eV}$, $0.10 \, \mathrm{eV}$, and $0.15 \, \mathrm{eV}$, taken from the same slice from the simulation volume. It can be seen that a neutrino follows the structure of CDM on large scales, while their small scale structure is suppressed by their velocity dispersions. This effect is more prominent for lighter neutrino masses. Similar to CDM, total matter overdensity remains visually indistinguishable across simulations as well; however, quantitative analysis can reveal up to 10\% variations in clustering strength, confirming that neutrinos modulate both their own, but also total clustering dynamics.

\begin{figure}[H]
    
    \includegraphics[width=0.5\linewidth]{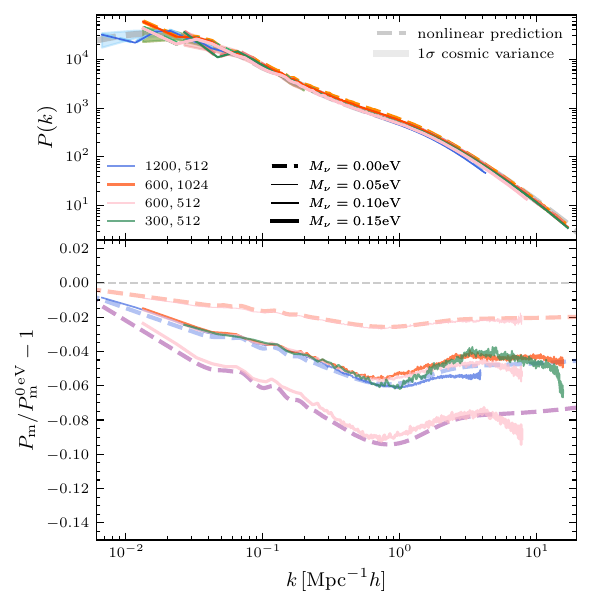}
    \caption{(\textbf{\hl{Top} %MDPI: Please ensure the different color of lines are corresponding with them explanation.
 Panel}): Simulated power spectra at redshift zero, color-coded by simulation configuration and line thickness corresponding to neutrino mass. Dashed lines denote nonlinear theoretical predictions, while shaded regions represent cosmological variance. (\textbf{Bottom Panel}): Residuals from neutrino-induced suppression of the power spectrum.}
    \label{fig:Pk}
\end{figure}

The power spectrum effectively quantifies the impact of neutrinos on the clustering of total matter. In Figure~\ref{fig:Pk}, we show the total matter power spectrum at redshift $z=0$ for all simulations and the power spectrum suppression for different neutrino masses. The results from simulations show that the suppression amplitudes are consistent with {\tt Halofit} predictions \cite{Halofit}, robust across resolutions and box sizes. 
Although from Figure~\ref{fig:rho}, the overdensities of more massive neutrinos are stronger, their greater fractional contribution to the total matter leads to a deeper total matter power suppression.

\subsection{Halo Statistics}
\begin{figure}[H]
   
    \includegraphics[width=0.5\linewidth]{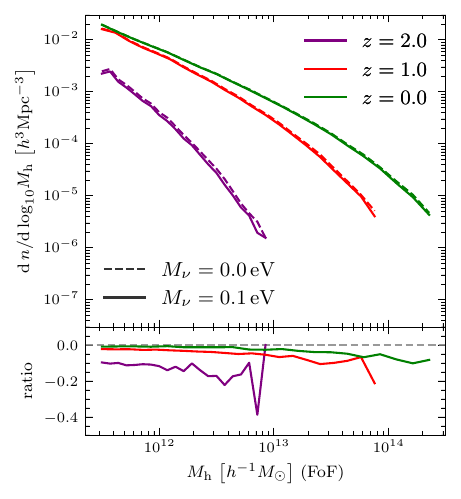}
    \caption{\hl{HMF} %MDPI: Please ensure the different color of lines are corresponding with them explanation.
 form neutrinos mass $M_\nu=0.1\,\rm eV$ (solid line) and  $M_\nu=0.0\,\rm eV$ (dashed line) in $z = 2$ (purple), $1$ (red) and $0$ (green). The ratios between different neutrino mass HMF (\mbox{$M_\nu=0.1\))}/HMF($M_\nu=0.0\)) are shown in the bottom panel.}
    \label{fig:HMF}
\end{figure}

The neutrino mass effects on halo statistics are shown in Figure~\ref{fig:HMF}. It compares two cosmological scenarios: one with massless neutrinos and the other with massive neutrinos of total mass $M_\nu=0.1 \rm ~eV$; furthermore, the evolution of the HMF from $z=2$ to $z=0$\linebreak is shown. 

The results reveal a suppression of the HMF in the presence of massive neutrinos compared to the massless neutrino case. This suppression is more pronounced for higher halo masses, as massive halos typically originate from high-density peaks in the initial matter distribution. The presence of massive neutrinos reduces the amplitude of these high-density peaks, thereby diminishing the formation of massive halos. Furthermore, the reduction is more substantial at higher redshifts, reflecting the cumulative impact of neutrinos on structure formation over cosmic time.

The suppression of the HMF due to massive neutrinos has important implications for observational probes such as galaxy clustering and weak gravitational lensing, as these observables are sensitive to the underlying halo population. Current and upcoming large scale structure surveys with higher precision, such as DESI and CSST, may provide constraints on neutrino masses by measuring the HMF at different redshifts.

\begin{figure}[H]
 
    \includegraphics[width=0.5\linewidth]{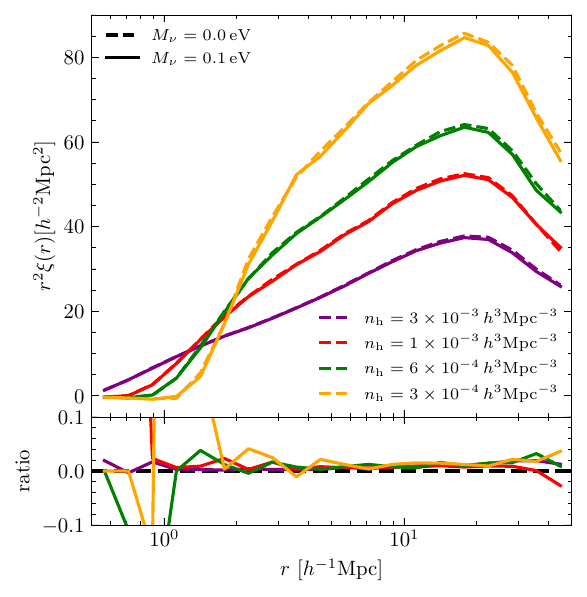}
    \caption{\hl{Two-point} %MDPI: Please ensure the different color of lines are corresponding with them explanation.
 correlation function of halos in different number density form the simulation with the sum of neutrino mass $M_\nu=0.1 \,\rm eV$ (solid line) and  $M_\nu=0.0 \, \rm eV$ (dashed line). The bottom panel shows the ratio between different neutrino masses.}
    \label{fig:2pcf}
\end{figure}

To ensure the consistency with observational studies that often select galaxy samples based on number density, we construct four halo samples with number densities of\linebreak $3 \times 10^{-3}\, h^{-3}\rm Mpc^3$, $1 \times 10^{-3}\, h^{-3}\rm Mpc^3$, $6 \times 10^{-4}\, h^{-3}\rm Mpc^3$, and $3 \times 10^{-4}\, h^{-3}\rm Mpc^3$ at redshift $z=0$. For each sample, we calculated the halo two-point correlation function, as shown in Figure~\ref{fig:2pcf}.

Interestingly, we find that the two-point correlation function in the massive neutrino cosmology is slightly higher (by a few percent) than the massless neutrino cosmology. This result appears counterintuitive, as neutrinos, due to their free-streaming behavior, are expected to suppress matter clustering and delay structure growth, as evidenced by the reduction in the matter power spectrum discussed in Section~\ref{MC}. The observed increase in the halo two-point correlation function can be attributed to the enhanced halo bias in the presence of massive neutrinos. Specifically, massive neutrinos increase the clustering strength of halos relative to the underlying matter distribution, which compensates for the suppression of the matter power spectrum and leads to a net increase in the halo two-point correlation function \cite{Euclid}.

% %%%%%%%%%%%%%%%%%%%%%%%%%%%%%%%%%%%%%%%%%%
% \section{Discussion}

% xxx

%%%%%%%%%%%%%%%%%%%%%%%%%%%%%%%%%%%%%%%%%%
\section{Conclusions}\label{sec4}
Building on the semi-linear method, we extend the algorithm to incorporate the influence of massive neutrinos on the expansion history of the universe and the evolution of their non-relativistic components, implementing these effects within a hierarchical multi-level PM grid structure, {\tt CUBE} 2.0. We further assume that neutrinos are completely homogeneous on small scales and introduced corresponding corrections to the PP \mbox{gravitational calculations}.

To enhance the efficiency of matter power spectrum estimation, we develop a segmented power spectrum estimation algorithm. By taking advantage of the intrinsic grid hierarchy of the multi-level PM scheme, this method introduces negligible additional memory or computational costs. Correction functions at each simulation step are obtained through interpolation/extrapolation of the power spectrum.

By a series of cosmological simulations featuring a range of neutrino masses, resolutions, and box sizes, the incorporation of massive neutrinos incurs minimal additional computational costs. The resulting matter power spectra and suppression features not only affirm the accuracy of our simulations but also highlight the robustness of the \mbox{numerical framework}.

The evolution of the HMF with redshift shows that neutrinos suppress halo formation, and this suppression gradually weakens at lower redshifts. This suggests that statistical analysis at different redshifts may offer a way to constrain the neutrino mass. Additionally, the two-point correlation function of halos indicates enhanced halo bias in massive \mbox{neutrino cosmologies.}

A key advantage of our cosmological simulation is its highly efficient memory usage for particle storage. While conventional codes like Gadget2 require approximately 80 bytes per particle \cite{Springel2005}, our implementation needs only 20 bytes per particle. Additionally, our neutrino simulation employs a power spectrum combining technique, eliminating the need to compute the global power spectrum and ensuring that neutrino-related calculations introduce negligible memory overhead. Thanks to these memory optimizations, our framework enables subsequent high-resolution simulations with a large number\linebreak of particles.

Leveraging the computational efficiency of our approach, future work will aim to carry out large-volume simulations to explore neutrino-induced parameter degeneracies and their interplay with other cosmological observables.

%%%%%%%%%%%%%%%%%%%%%%%%%%%%%%%%%%%%%%%%%%
%% optional
%\supplementary{The following supporting information can be downloaded at:  \linksupplementary{s1}, Figure S1: title; Table S1: title; Video S1: title.}

% Only for journal Methods and Protocols:
% If you wish to submit a video article, please do so with any other supplementary material.
% \supplementary{The following supporting information can be downloaded at: \linksupplementary{s1}, Figure S1: title; Table S1: title; Video S1: title. A supporting video article is available at doi: link.}

% Only used for preprtints:
% \supplementary{The following supporting information can be downloaded at the website of this paper posted on \href{https://www.preprints.org/}{Preprints.org}.}

% Only for journal Hardware:
% If you wish to submit a video article, please do so with any other supplementary material.
% \supplementary{The following supporting information can be downloaded at: \linksupplementary{s1}, Figure S1: title; Table S1: title; Video S1: title.\vspace{6pt}\\
%\begin{tabularx}{\textwidth}{lll}
%\toprule
%\textbf{Name} & \textbf{Type} & \textbf{Description} \\
%\midrule
%S1 & Python script (.py) & Script of python source code used in XX \\
%S2 & Text (.txt) & Script of modelling code used to make Figure X \\
%S3 & Text (.txt) & Raw data from experiment X \\
%S4 & Video (.mp4) & Video demonstrating the hardware in use \\
%... & ... & ... \\
%\bottomrule
%\end{tabularx}
%}
\vspace{6pt}
%%%%%%%%%%%%%%%%%%%%%%%%%%%%%%%%%%%%%%%%%%
\authorcontributions{\hl{Neutrino} %MDPI: Please add J.-h.H. in this part
 simulation code, power spectrum estimation, simulation testing and validation, power spectrum analysis B.-H.C.; prerequisite verification, halo statistical analysis J.-J.Z.; theoretical and methodological guidance H.-R.Y., Y.L., J.-H.H., Y.J.; {\tt CUBE} 2.0 code development H.-R.Y. All authors have read and agreed to the published version of the manuscript.}

\funding{This work is supported by the National Natural Science Foundation of China (NSFC) grant No. 12173030, 12303005, 12475058. Y.J.  is supported by NSFC 12133006, by National Key R\&D Program of China (2023YFA1607800, 2023YFA1607801),  and by 111 project No. B20019. Yu Liu acknowledges the Shuimu Tsinghua Scholar Program (No. 2022SM173). This work made use of the Gravity Supercomputer at the Department of Astronomy, Shanghai Jiao Tong University.}

\dataavailability{The original contributions presented in this study are included in the article. Further inquiries can be directed to the corresponding author.}%MDPI: {Please refer to suggested Data Availability Statements in section “MDPI Research Data Policies” at \href{https://www.mdpi.com/ethics}{https://www.mdpi.com/ethics}}.  In this section, please provide details regarding where data supporting reported results can be found, including links to publicly archived datasets analyzed or generated during the study. Please refer to suggested Data Availability Statements in section “MDPI Research Data Policies” at https://www.mdpi.com/ethics. You might choose to exclude this statement if the study did not report any data.

\conflictsofinterest{The authors declare no conflicts of interest.} 

%%%%%%%%%%%%%%%%%%%%%%%%%%%%%%%%%%%%%%%%%%
%% Optional

%% Only for journal Encyclopedia
%\entrylink{The Link to this entry published on the encyclopedia platform.}

% \abbreviations{Abbreviations}{
% The following abbreviations are used in this manuscript:
% \\

% \noindent 
% \begin{tabular}{@{}ll}
% MDPI & Multidisciplinary Digital Publishing Institute\\
% DOAJ & Directory of open access journals\\
% TLA & Three letter acronym\\
% LD & Linear dichroism
% \end{tabular}
% }

% %%%%%%%%%%%%%%%%%%%%%%%%%%%%%%%%%%%%%%%%%%
% %% Optional
% \appendixtitles{no} % Leave argument "no" if all appendix headings stay EMPTY (then no dot is printed after "Appendix A"). If the appendix sections contain a heading then change the argument to "yes".
% \appendixstart
% \appendix
% \section[\appendixname~\thesection]{}
% \subsection[\appendixname~\thesubsection]{}
% xxx

% \section[\appendixname~\thesection]{}
% xxx

%%%%%%%%%%%%%%%%%%%%%%%%%%%%%%%%%%%%%%%%%%
%\isPreprints{}{% This command is only used for ``preprints''.
\begin{adjustwidth}{-\extralength}{0cm}
%} % If the paper is ``preprints'', please uncomment this parenthesis.
%\printendnotes[custom] % Un-comment to print a list of endnotes

\reftitle{References}
\PublishersNote{}
%\isPreprints{}{% This command is only used for ``preprints''.
\end{adjustwidth}
%} % If the paper is ``preprints'', please uncomment this parenthesis.
\end{document}